\documentclass[a4paper]{article}
\usepackage{Odyssey2022}
\usepackage{epsfig,amssymb,amsmath,float,multirow,makecell}
\usepackage[flushleft]{threeparttable}
\usepackage{hyperref}
\usepackage{colortbl}
\hypersetup{colorlinks=true}

\ninept

\renewcommand\vec{\mathbf}

\title{Parameter-Free Attentive Scoring for Speaker Verification}

\makeatletter
\def\name#1{\gdef\@name{#1\\}}
\makeatother
\name{{\em Jason Pelecanos,  Quan Wang, Yiling Huang, Ignacio Lopez Moreno}
}

\address{Google LLC, USA \\
{\small \tt \{pelecanos, quanw, yilinghuang, elnota\}@google.com} }

\begin{document}
\maketitle

\begin{abstract}
This paper presents a novel study of parameter-free attentive scoring for speaker verification. Parameter-free scoring provides the flexibility of comparing speaker representations without the need of an accompanying parametric scoring model. Inspired by the attention component in Transformer neural networks, we propose a variant of the scaled dot product attention mechanism to compare enrollment and test segment representations. In addition, this work explores the effect on performance of (i) different types of normalization, (ii) independent versus tied query/key estimation, (iii) varying the number of key-value pairs and (iv) pooling multiple enrollment utterance statistics. Experimental results for a 4 task average show that a simple parameter-free attentive scoring mechanism can improve the average EER by 10\% over the best cosine similarity baseline.
\end{abstract}

\section{Introduction}

A trend in speaker recognition has focused on building systems utilizing fixed-size, compact, speaker embeddings. Examples include i-vectors~\cite{dehak_2011_1}, x-vectors~\cite{snyder_2018_1}, d-vectors~\cite{heigold_2016_1,wan_2018_1}, residual network variants~\cite{desplanques_2020_1} and adapted network representations~\cite{tan2021_1}. These embeddings are used directly for cosine based scoring as a similarity measure, or are included as part of a backend. The backend is trained either separately or jointly with the embedding network. For example, a PLDA model~\cite{prince_2007_1} was trained separately as part of a backend to score x-vectors~\cite{snyder_2018_1}. For contrast, there is an example of joint training of an embedding extractor with a PLDA backend~\cite{silnova_2020_1} or a decision network~\cite{pelecanos2021_1}.

With the introduction of neural network Transformers~\cite{vaswani2017_1}, the speaker recognition  community has considered how this work may be leveraged~\cite{tong_2020_1, safari_2020_1, mary_2020_1, wang_2022_1}. Typically, the transformer encoder is involved in generating a speaker embedding. A key element of the Transformer model is the attention mechanism. In this work, we are interested in how attention can be used to compare utterance embeddings.

Interestingly, some work previously examined how attention could be applied as part of backend in an end-to-end trained system. For example, in~\cite{Li2020TextIndependentSV}, the authors utilize a mutual-attention approach to perform a weighted combination of enrollment frames given an intermediate test utterance embedding and vice versa. To compute the speaker similarity, rather than just using speaker embeddings, both the enrollment and test utterances are required. For the work in~\cite{Jung2021graphatt}, the authors extract embeddings from short audio chunks and then use stacked modules of attention modeling to compute the final similarity score. Both approaches require scoring parameters to be estimated as part of the training process.
 
From an engineering perspective, having one single neural network that computes the speaker embedding without the need of additional parameters or models to compute the scores simplifies the deployment process of a speaker recognition system. It can also reduce risk related to version control~\cite{wang2020version} by avoiding the need to handle separate embedding generation and scoring networks.

Inspired by the Transformer work~\cite{vaswani2017_1}, we propose the use of a parameter-free attentive scoring mechanism\footnote{An open source implementation of attentive scoring and attentive temporal pooling modules based on Lingvo~\cite{shen2019lingvo} is provided at: \url{https://github.com/google/speaker-id/tree/master/lingvo}}. To the best of our knowledge, this is the first paper to perform attentive scoring using a parameter-free comparison of speaker embeddings. In contrast to~\cite{Li2020TextIndependentSV}, this approach does not need the original recordings once the speaker embeddings are generated. In addition, we perform a study of how performance changes with different types of feature normalization (Layer norm~\cite{ba2016_1}, L2-Norm, and no-norm) and modeling complexity (\emph{i.e.} the number of keys). We also evaluate scenarios with independent and tied queries/keys and as well as an approach for handling many enrollment utterances.

The remainder of the paper is as follows. Section~\ref{sec:attention_scoring} introduces the specifics of the parameter-free attentive scoring work; Section~\ref{sec:system_description} provides details on the broader system used in the analysis; Section~\ref{sec:experiment_and_results} follows this with an experimental analysis; and Section~\ref{sec:conclusion} wraps up with the conclusion.

\section{Parameter-free attentive scoring}
\label{sec:attention_scoring}

\subsection{Core approach}

In this section, we detail an attentive scoring mechanism based on the work of~\cite{vaswani2017_1}. First, we assume that a jointly trained embedding network will generate the fixed dimensional speaker representations for both enrollment and test utterances. A test utterance representation $\vec{U}_{t}$ can be composed of $M$ \textit{query} vectors and $M$ corresponding \textit{value} vectors $\vec{U}_{t}=\{\vec{q}_m, \vec{t}_m\}_{m=1,\ldots,M}$. For each enrollment utterance, there are $M$ \textit{key} vectors and $M$ corresponding \textit{value} vectors. Given $E$ enrollment utterances for a speaker, there are $N$ \textit{key} vectors (where $N=M \times E$) and $N$ corresponding \textit{value} vectors given by $\vec{U}_{e}=\{\vec{k}_n, \vec{e}_n\}_{n=1, \ldots, N}$. Keys and queries are of dimension $d_k$, while values are of dimension $d_v$. Let a parameter $\alpha$ represent a scaling factor (or \textit{temperature} parameter~\cite{Hinton2015DistillingTK}) for the softmax function; which can be jointly trained or directly specified as a hyperparameter (for example~\cite{vaswani2017_1} set $\alpha = 1 / \sqrt{d_k}$). The parameter-free (apart from $\alpha$) attentive score $s_{\text{att}}$ may be calculated as follows:

\begin{align}
\label{eqn:attention}
s_{\text{att}}(\vec{U}_{t}, \vec{U}_{e}) &= \sum_{m=1}^{M} \sum_{n=1}^{N} w_{mn} \vec{t}_m \cdot \vec{e}_n \\
\label{eq:softmax}
w_{mn} &= \frac{\exp (\alpha \, \vec{q}_m \cdot \vec{k}_n)}{\sum_{i=1}^M \sum_{j=1}^N \exp ( \alpha \, \vec{q}_i \cdot \vec{k}_j)}
\end{align}
\vspace{0.2cm}

Here ($\cdot$) represents the dot product.

The scoring function can be interpreted simply as a weighted combination of dot products across embeddings (\emph{i.e.} the value vectors), where the weights are determined using the softmax calculation (based on query and key comparisons). In addition, the softmax weight function enables \textit{value} vectors to be compared across multiple key/query indexes with an emphasis on the most relevant pairings. We found that normalization is important for enhancing performance and we discuss this next.

\subsection{Normalization}

In this work we examine different types of normalization and evaluate their effectiveness in experiments section.

\subsubsection{Layer normalization}
One approach is Layer Normalization~\cite{ba2016_1} where the features across a single example are transformed to have a zero mean and unit standard deviation. This is followed by the application of a per element bias and gain. While there are other possible configurations, we apply Layer Normalization at the utterance representation level.

\subsubsection{Query, key and value normalization}
In Equations~\ref{eqn:attention} and~\ref{eq:softmax}, the key and value vectors may be used as is or they can each be normalized. One approach we assess is to apply L2 length normalization to the query, key and value vectors. Note that if the value vectors are L2 length normalized, the attentive scoring function effectively calculates a weighted combination of cosine similarity scores.

\subsubsection{Global length normalization}
\label{sec:global_norm_variant}
Equation~\ref{eqn:attention} can be interpreted as one large dot product between enrollment and test value vector representations with $\{w_{mn}\}_{\forall m,n}$ split between enrollment and test. The score $s_{\text{att}}(\vec{U}_{t}, \vec{U}_{e})$ can be calculated as follows.
\begin{align}
\label{eqn:attentiondotprod}
s_{\text{att}}(\vec{U}_{t}, \vec{U}_{e}) &= \vec{a} \cdot \vec{b}
\end{align}
\begin{align}
\vec{a} = \left(\begin{smallmatrix}\left(\begin{smallmatrix}
\sqrt{w_{11}} \vec{t}_1 \\
\vdots \\   \\
\sqrt{w_{1N}} \vec{t}_1
\end{smallmatrix}\right)\\
\vdots \vspace{0.15cm} \\
\left(\begin{smallmatrix}
\sqrt{w_{M1}} \vec{t}_M \\
\vdots  \\ \\
\sqrt{w_{MN}} \vec{t}_M
\end{smallmatrix}\right)
\end{smallmatrix}\right)
\hspace{0.5cm}\vec{b}= \left(\begin{smallmatrix}\left(\begin{smallmatrix}
\sqrt{w_{11}} \vec{e}_1 \\
\vdots \\ \\
\sqrt{w_{1N}} \vec{e}_N
\end{smallmatrix}\right)\\
 \vdots \vspace{0.15cm} \\
\left(\begin{smallmatrix}
\sqrt{w_{M1}} \vec{e}_1 \\
\vdots \\ \\
\sqrt{w_{MN}} \vec{e}_N
\end{smallmatrix}\right)
\end{smallmatrix}\right)
\end{align}
\vspace{0.2cm}

We can apply L2 length normalization to these representations to obtain a result that relates to a cosine or correlation calculation for the concatenated vector representations $\vec{a}$ and $\vec{b}$. The normalized score may be calculated as:
\begin{align}
\label{eqn:globalnormattention}
s_{\|\text{att}\|} (\vec{U}_{t}, \vec{U}_{e}) &= \frac{s_{\text{att}}(\vec{U}_{t}, \vec{U}_{e})} { \sqrt{\|\vec{a}\|_2^2 \; \|\vec{b}\|_2^2} } = \frac{\vec{a} \cdot \vec{b}} { \sqrt{\|\vec{a}\|_2^2 \; \|\vec{b}\|_2^2} }
\\
\|\vec{a}\|_2^2 &= \sum_{m=1}^{M}  \left( \sum_{n=1}^{N} w_{mn} \right) \|\vec{t}_m\|_2^2  \\ 
    \|\vec{b}\|_2^2 &= \sum_{n=1}^{N} \left(\sum_{m=1}^{M} w_{mn}\right)  \|\vec{e}_n\|_2^2
\end{align}

Here, $\|\vec{a}\|_2^2$ and $\|\vec{b}\|_2^2\,$ represent the squares of the Euclidean norms of the corresponding vectors $\vec{a}$ and $\vec{b}$. Similarly, $\|\vec{t}_m\|_2^2$ and $\|\vec{e}_n\|_2^2$ represent the squares of the Euclidean norms for $\vec{t}_m$ and $\vec{e}_n$. This normalized dot product is evaluated in the experiments as the \textit{Global L2-norm} system. 

\subsection{Implementation considerations}
This section covers the implementation considerations related to attentive scoring. 

\subsubsection{Generating keys, queries and values}
\begin{figure}[t]
\includegraphics[trim={3cm 0.8cm 4.2cm 0.7cm},clip,width=\columnwidth]{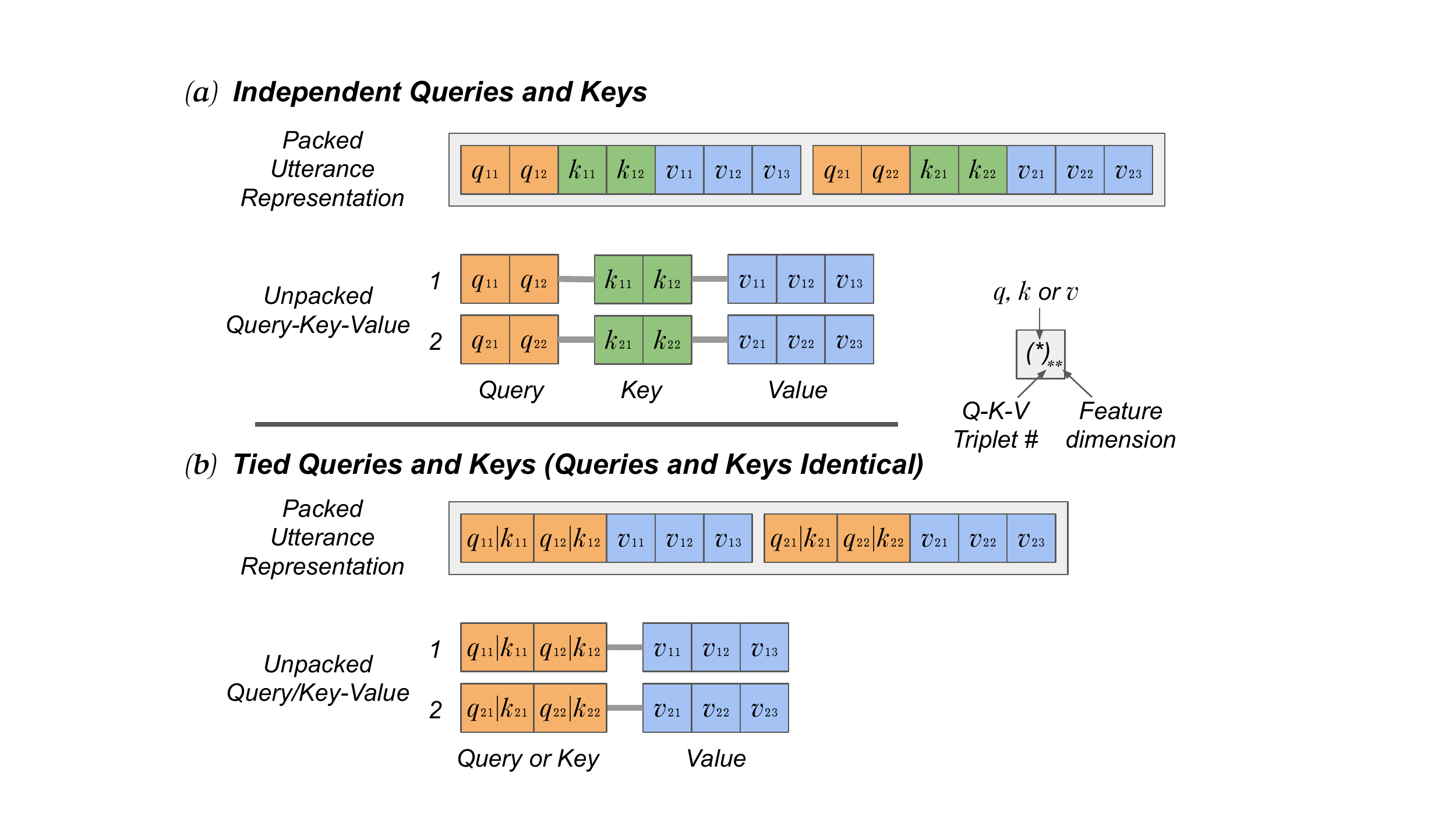}
\caption{{\it Packed and unpacked forms of an utterance representation for: (a) 2 sets of query-key-value triplets where the queries and keys are independently estimated. Note that for the current scoring paradigm the keys from test utterances and the queries from enrollment representations would not be used. (b) 2 sets of query-key-value outputs where queries and keys are tied and treated interchangeably. Here, as an example, $q_{12}|k_{12}$ means that the corresponding element in the vector can represent either a query term $q_{12}$ for test or a key term $k_{12}$ for enrollment. Additionally, $v_{12}$ represents the corresponding element from \textit{value} vectors from either $\vec{e}$ for enrollment or $\vec{t}$ for test. For both (a) and (b) the \textit{value} vectors are always tied for enrollment and test utterances.}}
\label{fig:packunpack}
\end{figure}

First we discuss how to generate the key/query/value vector representations from the speaker embedding network. A speaker embedding network based on Conformers~\cite{gulati2020conformer} is used to generate a fixed dimensional representation. This is followed by a linear layer (without activation) where the number of output nodes is equivalent to the number of parameters needed to estimate the key/query/value vectors. The relevant vectors are unpacked from the speaker representation generated by the embedding network. Figure~\ref{fig:packunpack}\hyperref[fig:packunpack]{{\it a}} shows how a packed speaker representation containing 2 query-key-value vectors is unpacked into independent queries (2 dimensions), keys (2 dimensions), and values (3 dimensions). Figure~\ref{fig:packunpack}\hyperref[fig:packunpack]{{\it b}} shows how a query-key-value vector is unpacked when the queries and keys have values that are tied (\textit{i.e.} they are the same).

There is an important clarification regarding enrollment utterances and the effective number of keys/queries/values. As presented in Figure~\ref{fig:packunpack}, the number of enrollment keys (and similarly queries and values) for a speaker is the number of enrollment utterances for the speaker times the number of enrollment keys per utterance. Since there is only one test utterance involved in a trial, this equates to $M$ keys/queries per utterance. In the experiments there are up to 6 enrollment utterances per speaker. So, for the case of 6 enrollment segments, the number of enrollment keys, $N$, would be equal to $6\times M$.

\subsubsection{Setting the softmax temperature}
One clarification is how the softmax scaling parameter alpha ($\alpha$) is set. In the attention paper~\cite{vaswani2017_1}, it is suggested that the softmax scaling parameter ($\alpha$ in Equation~\ref{eq:softmax}) can be specified manually as $1 / \sqrt{d_k}$. It is a scaling factor (or temperature) used to govern how much the highest score dominates the softmax function. In this work, we explicitly train the scaling factor which can be included with the output of the speaker embedding network if required.

\subsubsection{Handling many enrollment utterances}
Another consideration is the handling of many enrollment utterances, since an embedding is generated and kept for each utterance. Having many enrollment utterances can make the size of the effective speaker representation significantly larger. For $E$ enrollment utterances for a speaker, the total number of values to manage is $M \times E \times (d_k + d_v)$. In practical applications, we aim to keep this size reasonable. Some approaches to address this include hierarchical clustering, variability modeling and the most straightforward approach being a simple average of enrollment utterance representations. We examine the performance of simple averaging in the experiments in Section~\ref{sec:many_enroll}.

\section{System description}
\label{sec:system_description}

\begin{figure}[t]
\includegraphics[width=\columnwidth]{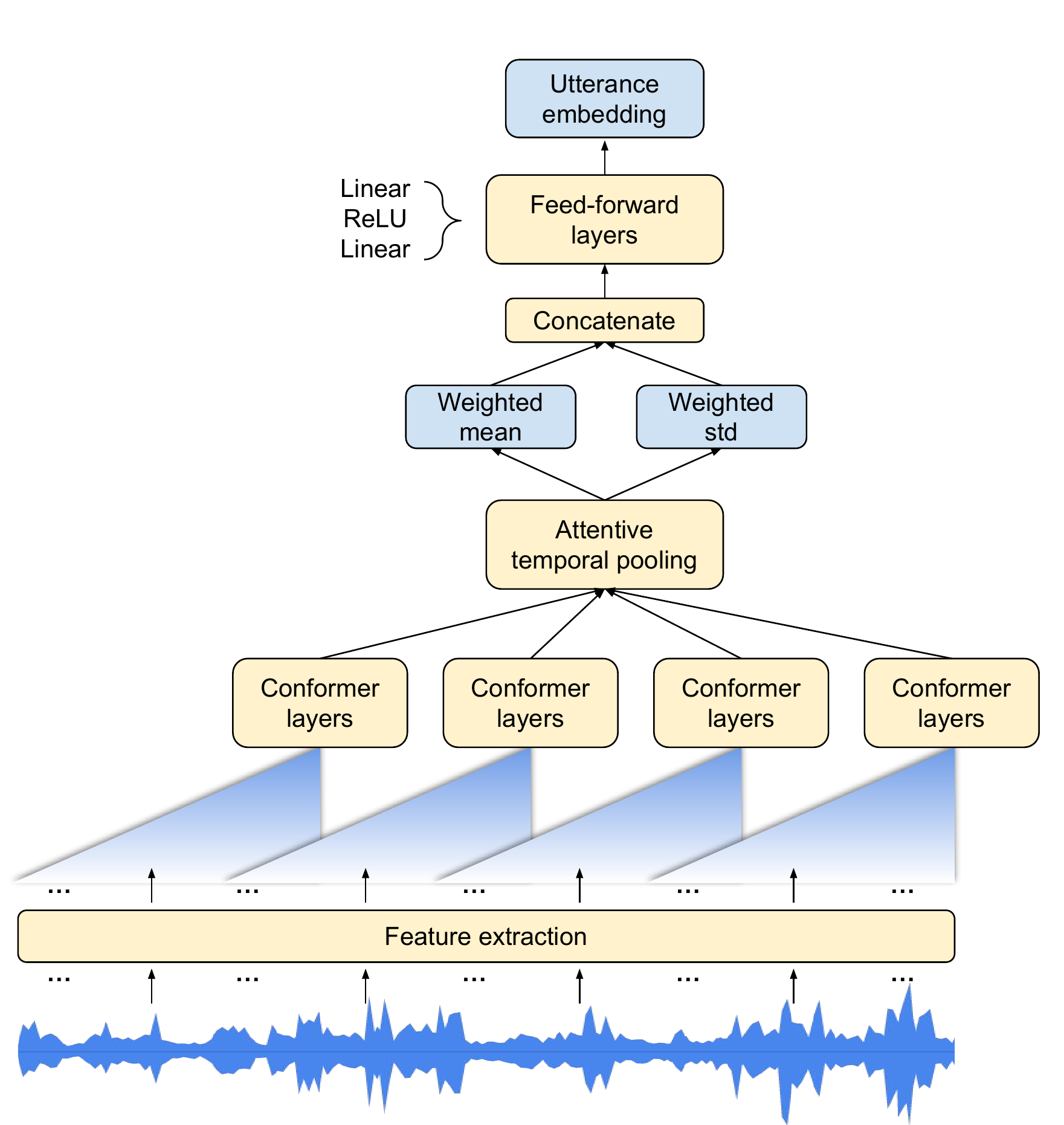}
\caption{{\it Diagram of the utterance embedding extractor. The extractor generates embeddings for both enrollment and test utterances.}}
\label{fig:system_description}
\end{figure}

This section covers a description of the system used to generate the utterance embeddings, details on the attentive scoring approach and information on system training. First we discuss how the utterance embeddings are generated (see Figure~\ref{fig:system_description}). 

The feature extraction front-end essentially consists of calculating log Mel-filterbank energy feature vectors which are then stacked. More specifically, given the speech signal, automatic gain control (AGC) is first applied~\cite{prabhavalkar2015automatic}. This is followed by framing into 32ms Hanning windows with a 10ms frame shift. For each frame, 128 log Mel-filterbank energies are calculated across a 125-7500Hz bandwidth. The final features consist of 4 stacked frames sampled every 3 frames. The resulting features are 512 dimensions with a 30ms frame shift.

These features are fed into a stack of conformer layers~\cite{gulati2020conformer}. Our system uses a stack of 12 conformer layers with a native dimensionality of 512. Each layer has a relative positional embedding dimension of 512 and the attention mechanisms consist of 4 heads. The convolutional component layers span 32 elements. The conformer network generates a 512 dimensional output for each input frame. Last but not least, we perform a stack by 2 with a stride of 2 after the third conformer layer, and insert a non-linear projection layer with an output dimension of 512 after the fourth conformer layer. To handle the `stack by 2' processing, we stream packets of 2 frames at inference time.

An attentive temporal pooling (weighted averaging) layer~\cite{wang2022attentive} is applied to the output of the conformer network to aggregate the first and second order statistics over the duration of the recording. The weighted mean and standard deviation statistics of the conformer outputs are calculated. The weight is a 0 to 1 value generated for each frame by taking a linear combination of the conformer outputs, adding a bias term, and passing it through a sigmoid non-linearity. These per-sample weights are used to generate the \textit{running} mean and standard deviation terms which are then concatenated to give 1024 dimensional features for each frame. In many situations, only the last frame of the model output is used, but intermediate statistics are available if they are needed for various streaming applications. Other works have considered related approaches~\cite{Chowdhury2018_1, zhu2018_1}.

Given the cumulative mean and standard deviation statistics at the final frame, the system applies an affine transformation followed by a ReLU~\cite{nair2010_1,Fukushima1980_1} non-linearity to give 512 output dimensions. This is followed by a linear transformation that has a range of output nodes depending on the number of required parameters. For example, for the cosine calculation, the numbers of output nodes tested were 256, 512 and although not necessary 2,304. For the attentive scoring approach, there were up to 36,864 output nodes.

The utterance embedding extractor is used to generate representations for both enrollment and test utterances. These representations are compared using the cosine similarity (as baseline) or the attentive scoring mechanism to generate a speaker similarity score. During training, these scores are scaled (using a scaling parameter that is trained) and are optimized according to a \textit{generalized end-to-end extended-set softmax} loss~\cite{pelecanos2021_1}. Optimization is performed across randomly generated mini-batches, where each mini-batch compares the statistics of 128 utterances from 16 speakers. More details are available here~\cite{pelecanos2021_1, wan_2018_1}. In this work, the system is trained with Adam optimization~\cite{Kingma2015_1} and uses a warm-up (ramp-up) schedule of 50k steps before allowing the learning rate to decrease until it reaches 500k training steps (a similar approach was used in~\cite{vaswani2017_1}).

\section{Experiments}
\label{sec:experiment_and_results}

In this section, we describe the experimental data and perform various experiments examining the effect of normalization, query/key configuration, the number of keys, and enrollment utterance consolidation.

\subsection{Training and evaluation data}

\begin{table}
\centering
 \begin{threeparttable}
  \caption{\label{tab:trainevaldata}  {\it Composition of the training and evaluation data across languages for the (mostly) vendor collected data. All numbers are indicated in thousands (as indicated by [k]).}}
\vspace{2mm}
\setlength{\tabcolsep}{3.2pt}
\begin{tabular}{|l| c c | c c c c|}
\hline
   & \multicolumn{2}{c|}{\textbf{Training}} & \multicolumn{4}{c|}{\textbf{Evaluation}} \\
    \multicolumn{1}{|c|}{\textbf{Language}}      & \textbf{Spk}  & \textbf{Utt} & \textbf{Spk}  & \textbf{Utt} & \textbf{Tar} & \textbf{Non} \\
     & [k]  & [k] & [k]  & [k] & [k] & [k] \\ \hline
  English (India)              & 6.0 & 2900 & 1.6 & 264 &  200 & 200 \\
  English (US)               & 46.7 & 3329 & 12.4 & 543 & 200 & 200 \\
  French              & 5.8 & 2458 & 1.4 & 161 &  152 & 200 \\ 
  Hindi              & 8.3 & 1642 & 2.4 & 106 &  93 & 200 \\
  Italian              & 5.2 & 2251 & 1.2 & 102 &  95 & 200 \\
  Japanese              & 6.2 & 2048 & 1.7 & 106 & 97 & 200 \\
  Korean               & 5.4 & 1407 & 1.5 & 160 & 151 & 200 \\
  Portuguese (Brazil)   & 5.8 & 1675 & 1.5 & 107 & 97 & 200 \\
  Spanish                   & 4.6 & 2189 & 1.2 & 113 &  106 & 200 \\
  \emph{Other Data/Sources$^\dagger$} & 139.8 & 54499 & -    &  -    &   -   &   -   \\ \hline
\end{tabular}
\begin{tablenotes}
            \item[$\dagger$] {\it  Includes vendor collected data from languages outside of the languages mentioned as well as LibriVox, CN-Celeb, and LDC sourced data.}
\end{tablenotes}
\end{threeparttable}
\end{table}

The training set\footnote{A training set of 72 languages: Afrikaans, Akan, Albanian, Arabic, Assamese, Basque, Bengali, Bulgarian, Cantonese, Catalan, Croatian, Czech, Danish, Dutch, English, Estonian, Filipino, Finnish, French, Galician, German, Greek, Gujarati, Haitian, Hausa, Hebrew, Hindi, Hungarian, Icelandic, Igbo, Indonesian, Italian, Japanese, Kannada, Kazakh, Kinyarwanda, Korean, Lithuanian, Macedonian, Malagasy, Malay, Malayalam, Mandarin Chinese (Simplified), Mandarin Chinese (Traditional), Marathi, Mongolian, Norwegian, Odia, Oromo, Polish, Portuguese, Punjabi, Romanian, Russian, Samoan, Serbian, Sesotho, Sindhi, Slovak, Spanish, Swedish, Tamil, Telugu, Thai, Tibetan, Turkish, Ukrainian, Urdu, Uzbek, Vietnamese, Yoruba, Zulu.} in Table~\ref{tab:trainevaldata} consists mostly of vendor collected speech queries from different language varieties using devices such as laptops and cell phones. It also includes LibriVox~\cite{librivox_2020_1}, CN-Celeb~\cite{fan_2020_1} and LDC sourced data (Fisher~\cite{Cieri2004_1}, Mixer 4/5~\cite{Brandschain2020_1}, TIMIT~\cite{garofolo_1993_1}).

For system training, we apply data augmentation techniques based on noise and room simulation effects~\cite{lippmann1987multi,ko2017study,kim2017generation}. Similar augmentation techniques~\cite{garcia-romero_2012_1,lei_2012_1,avila_2014_1,snyder_2018_1,huang_2019_1} were previously used for speaker recognition. Noise is added to the training utterances with an SNR ranging from 3dB to 15dB. The signal amplitude is also varied from a scale of 0.01x to 1.2x.

Performance is evaluated by averaging the Equal Error Rate (EER) across the 9 language varieties listed in Table~\ref{tab:trainevaldata}. We evaluate performance across two dimensions. The first is whether speaker enrollment consists of a single utterance or multiple (up to 6) utterances. For the single utterance enrollment task, we select the same trials as in the multiple enrollment utterance task, except that only the first enrollment utterance is chosen for enrollment. It is expected that the multiple enrollment case should have a significantly lower EER than the single enrollment result. The second factor considers if the speech is clean (the original data) or noisy. The noisy data is simply the clean test segment data (not the enrollment segments) with previously \textit{unseen} noise (at 3-15dB SNR) and reverberation effects applied.

\subsection{Embedding normalization approaches}

Table~\ref{tab:baseline_embed_norm} compares the cosine similarity baseline systems with different attentive scoring configurations with tied keys and queries. Empirically, we found that the best cosine similarity result was achieved by calculating the cosine similarity between the test utterance representation and the average of the L2 normalized enrollment utterance representations. For cosine similarity scoring we compare three results. The first has an output embedding size of 256. We also include the result for an embedding size of 512, which is the maximum appropriate embedding size given a native conformer model dimension of 512. Unlike attentive scoring (which is a non-linear scoring function), there should be no additional benefit to increasing dimensionality past the native dimensionality when using a linear transform followed by cosine similarity. This is supported by the third row of cosine results which has an output dimension of 2,304. Across the 3 baseline systems, apart from the $2.31$\% EER result, all other numbers are comparable and have a very similar task average.

We also compare the baseline systems to attention models with different query/key/value normalization schemes. The first attention result is without normalization (\textit{i.e.} just using the raw linear layer outputs). The next result considers the use of Layer Normalization~\cite{ba2016_1}. In this implementation, we apply layer normalization on a per utterance representation basis. The results may be different if normalization was applied on a per-key and per-value vector basis. The next two rows of results (\textit{Key \& Value L2-Norm}) explore the utility of applying L2 length normalization of both queries/keys and values. There is some improvement observed for both the 8 and 32 key-value pairs over previous rows of results. However, even better results (last two rows, \textit{Key \& Global L2-Norm}) are achieved when each key vector is first L2-normalized followed by applying \textit{Global} L2-normalization (see Section~\ref{sec:global_norm_variant} for details). Of these two sets of results, the best result is for 32 key-value pairs and represents a 10\% relative improvement in average task EER over the best cosine similarity baseline.

\begin{table*}[t]
\caption{\label{tab:baseline_embed_norm} {\it A comparison of cosine similarity baseline systems and attentive scoring normalization approaches. The attentive scoring results shared here are for tied query-key vectors. Models that generate the speaker representations vary from using 21.4M parameters (for the 256 dimensional embedding cosine similarity system) to 22.4M parameters (for systems with 2,304 total dimensions). For tied query-key attentive scoring, the total output dimension is calculated as the sum of the per key and per value dimensions all multiplied by the number of key-value pairs. For the system marked with an (A), the total output dimensionality per utterance is $2304 = (32+256) \times 8$. In the table we mark two systems with (A) and (B) and they will be used in latter comparisons.}}
\vspace{2mm}
\centerline{
\setlength\tabcolsep{4pt}
\begin{tabular}{|l|c|ccc|cc|cc|c|}
\hline
  & \multirowcell{3}{\# \\ Key-Value \\ Pairs} & \multicolumn{3}{c|}{\# Dimensions} & \multicolumn{2}{c|}{Single Enroll} & \multicolumn{2}{c|}{Multi Enroll} & Task \\
 \multicolumn{1}{|c|}{System}   &  & Per & Per &  & \multicolumn{2}{c|}{EER (\%)} & \multicolumn{2}{c|}{EER (\%)} & Average \\
 &       & Key & Value & Total &\multicolumn{1}{c}{Clean} & Noisy & \multicolumn{1}{c}{Clean} & Noisy & EER (\%) \\
\hline  \hline
\hspace{0.15cm}\multirow{3}{*}{Baseline (Cosine Similarity)} & - & - & -  & 256 & 2.23 & 3.38 & 0.67 & 1.23 & 1.88 \\
         & - & - & - & 512 &  2.31 & 3.34 & 0.68 & 1.20 & 1.88 \\
         & - & - & - & 2304 &  2.22 & 3.35 & 0.67 & 1.23 & 1.87\\ \hline
Attentive Scoring: &  & & &  &  &  & & & \\
\hspace{0.1cm}No normalization & 8 & 32 & 256 & 2304 & 2.15 & 3.26 & 0.69 & 1.29 & 1.85 \\
\hspace{0.1cm}Layer Normalization~\cite{ba2016_1} & 8 & 32 & 224 & 2048 & 2.15 & 3.30 & 0.71 & 1.30 & 1.86 \\
\hspace{0.1cm}Key \& Value L2-Norm (A) & 8 & 32 & 256 & 2304 & 2.03 & 3.20 & 0.65 & 1.21 & 1.77 \\
\hspace{0.1cm}Key \& Value L2-Norm & 32 & 16 & 48 & 2048 & 2.01 & 3.20 & 0.65 & 1.22 & 1.77 \\
\hspace{0.1cm}Key \& Global L2-Norm & 8 & 32 & 256 & 2304 & 1.94 & 3.15 & 0.61 & 1.20 & 1.72 \\
\hspace{0.1cm}Key \& Global L2-Norm (B) & 32 & 16 & 48 & 2048 & \textbf{1.93} & \textbf{3.02} & \textbf{0.60} & \textbf{1.15} & \textbf{1.68} \\
\hline
\end{tabular}}
\end{table*}

\subsection{Tied versus independent query-key estimation}

In this section (see Table~\ref{tab:key_query_comparison}) we compare performance differences between a system where the queries and keys are identical and a system where queries are estimated independently to keys. The results are mixed and may be dependent on the particular configuration of the queries-keys-values (\textit{i.e.} fewer/more queries or keys). For system \textit{Key \& Value L2-Norm (A)}, with only 8 queries/keys, independent query and key estimation provides improvement in 3 of the 4 base cases. For the \textit{Key \& Global L2-Norm (B)} system, with significantly more queries/keys at 32, tying the queries and keys is consistently better. Having independent queries and keys allows the model to set each query and its corresponding key to be very different and results in increased flexibility in the final scoring function. The attentive scoring function does not have this flexibility for tied queries and keys. It is forced to give high scores to speaker representations that are the same even though they may be derived from recordings with only noise.

\begin{table}[t]
\caption{\label{tab:key_query_comparison} {\it Table comparing the results of the tied versus the independently estimated query and key vector approaches. The two systems listed (``Key \& Value L2-Norm (A)" and ``Key \& Global L2-Norm (B)") are the same as the corresponding ones in Table~\ref{tab:baseline_embed_norm} for the tied query-key case. For independent queries and keys, additional speaker representation parameters are estimated to allow for the queries and keys to be independent. For example, the ``Key \& Value L2-Norm (A)" (query-key tied) system, with $8$ keys, $32$ dimensions per key and $256$ dimensions per value vector, would require $8 \times 32 = 256$ additional parameters to support independent queries and keys. In this case, the number of parameters is increased from 2,304 to 2,560.}}
\vspace{2mm}
\centerline{
\setlength\tabcolsep{2.3pt}
\begin{tabular}{|c|c|c|cc|cc|c|}
\hline
  & \multirowcell{3}{Query\\\&\\Key} & \multirowcell{3}{Total\\Dims} & \multicolumn{2}{c|}{Single Enroll} & \multicolumn{2}{c|}{Multi Enroll} & Task \\
 \multicolumn{1}{|c|}{System}   & &  & \multicolumn{2}{c|}{EER (\%)} & \multicolumn{2}{c|}{EER (\%)} & Average \\
 &      &  & \multicolumn{1}{c}{Clean} & Noisy & \multicolumn{1}{c}{Clean} & Noisy & EER (\%) \\
\hline  \hline
\multirowcell{2}{(A)} & Tied & 2304 & 2.03 & 3.20 & 0.65 & 1.21 & 1.77 \\
      & Indep & 2560 & 2.06 & 3.08 & 0.63 & \textbf{1.14} & 1.73 \\ \hline
\multirowcell{2}{(B)} & Tied & 2048 & \textbf{1.93} & \textbf{3.02} & \textbf{0.60} & 1.15 & \textbf{1.68} \\
  & Indep & 2560 & 2.04 & 3.12 & 0.62 & 1.17 & 1.74 \\
\hline
\end{tabular}}
\end{table}

\subsection{Varying the number of keys}

It is also helpful to understand performance across different numbers of keys. Table~\ref{tab:function_of_num_keys} shares performance numbers as the number of keys is increased from 1 and doubled up to 128 keys. With our non-linear scoring function, we may be able to improve performance by increasing the dimensionality of the output representation beyond the native model dimensionality of 512. As highlighted earlier, this would not be the case for the regular cosine similarity. In examining the results, the performance improves as the number of keys is increased. The best (overall) task average result is reached at 64 keys.

\subsection{Managing many enrollment utterances}
\label{sec:many_enroll}
For the proposed attentive scoring models, one embedding is generated for each utterance in enrollment. For the case where only a few utterances are involved, this is a non-issue. For situations where there is a large number of utterances for speaker enrollment (such as when adaptation is allowed), there is the practical consideration of storing many enrollment representations for each speaker. There are several approaches to this problem. A simple approach is to average the utterance embeddings. In these experiments we average before applying further processing such as normalization. Table~\ref{tab:multi_enroll} presents three sets of results. The first row of results relates to the \textit{Key \& Value L2-Norm (A)} result (from Table~\ref{tab:baseline_embed_norm}). This represents the regular attentive scoring system where all enrollment utterances are used jointly in both system training and evaluation. The second row of results is determined by using the same system as before except during evaluation the system uses the mean of the enrollment utterances and considers it as a single enrollment representation. The last row of results involves the mean being calculated for both system training and evaluation. Given that all systems are trained using multiple enrollment utterances for a speaker, it is interesting to note that either enrollment utterance averaging approach is reasonable for the \textit{Multi Enroll} tasks. However, if averaging is done as part of training, a noteworthy performance decrease is observed for the \textit{Single Enroll} tasks.

\begin{table}
\centering
\begin{threeparttable}
\caption{\label{tab:function_of_num_keys} {\it Table of EERs as a function of the number of key-value pairs. These results are based on the ``Key \& Value L2-Norm (A)'' system from Table~\ref{tab:baseline_embed_norm} except that the number of keys is varied. (For reference, this system has tied queries/keys, 32 key dimensions and 256 value dimensions.)}}
\vspace{2mm}
\setlength\tabcolsep{3.5pt}
\begin{tabular}{|c|c|cc|cc|c|}
\hline
   \# Key-& \multirowcell{3}{Total\\Dims} & \multicolumn{2}{c|}{Single Enroll} & \multicolumn{2}{c|}{Multi Enroll} & Task \\
   Value &  & \multicolumn{2}{c|}{EER (\%)} & \multicolumn{2}{c|}{EER (\%)} & Average \\
 Pairs &       & \multicolumn{1}{c}{Clean} & Noisy & \multicolumn{1}{c}{Clean} & Noisy & EER (\%)\\
\hline  \hline
1 & 288 & 2.17 & 3.20 & 0.75 & 1.31 & 1.86 \\
2 & 576 & 2.08 & 3.23 & 0.70 & 1.30 & 1.83 \\
4 & 1152 & 2.12 & 3.21 & 0.68 & 1.21 & 1.80 \\
8 & 2304 & 2.03 & 3.20 & 0.65 & 1.21 & 1.77 \\
16 & 4608 & 1.99 & \textbf{3.09} & 0.62 & \textbf{1.15} & 1.71 \\
32 & 9216 & 1.95 & 3.10 & 0.61 & 1.16 & 1.70 \\
64 & 18432 & \textbf{1.89} & 3.10 & \textbf{0.60} & \textbf{1.15} & \textbf{1.68} \\
128 & 36864 & 1.97 & 3.10 & 0.61 & 1.18 & 1.72 \\
\hline
\end{tabular}
\end{threeparttable}
\end{table}

\begin{table}
\caption{\label{tab:multi_enroll} {\it A comparison of approaches addressing the multiple enrollment utterances for the ``Key \& Value L2-Norm (A)" system.} }
\vspace{2mm}
\centerline{
\setlength\tabcolsep{3.5pt}
\begin{tabular}{|cc|cc|cc|c|}
\hline
 \multicolumn{2}{|c|}{Estimation}  & \multicolumn{2}{c|}{Single Enroll} & \multicolumn{2}{c|}{Multi Enroll} & Task \\
 \multicolumn{2}{|c|}{Method}    & \multicolumn{2}{c|}{EER (\%)} & \multicolumn{2}{c|}{EER (\%)} & Average \\
 Train & Eval  &  \multicolumn{1}{c}{Clean} & Noisy & \multicolumn{1}{c}{Clean} & Noisy & EER (\%) \\
\hline  \hline
Joint & Joint  & \textbf{2.03} & \textbf{3.20} & \textbf{0.65} & 1.21 & 1.77 \\
Joint & Mean & \textbf{2.03} & \textbf{3.20} & \textbf{0.65} & 1.16 & \textbf{1.76} \\ Mean & Mean & 2.32 & 3.57 & 0.67 & \textbf{1.14} & 1.92 \\
\hline
\end{tabular}}
\end{table}

\section{Conclusion}
\label{sec:conclusion}
We proposed a parameter-free attentive scoring approach to meet the objectives of improving performance while using a relatively simple (parameter-free) scoring mechanism. We evaluated different normalization techniques and results suggest that applying per query/key L2-normalization followed by global L2-normalization was the most effective. We also observed that using independently estimated queries and keys (versus tied queries and keys) gave mixed results. We note that the approach may be helpful depending on the evaluation data and the attentive scoring configuration. Furthermore, we evaluated the key and value L2-norm system across different numbers of keys and found that its best task average EER was at 64 keys. This type of system was also shown to work well when scoring trials using a single average representation of multiple enrollment utterances.

Future work could consider other vector normalization techniques and query-key-vector configurations. Another path is to consider better approaches for converting from frame based network outputs to the final fixed dimensional representations. Currently we capture (single-head) attention based mean and standard deviation statistics. However, frame-based pooling using multi-headed attention could be an appropriate extension.

\bibliographystyle{IEEEbib}
\bibliography{Odyssey2022_BibEntries}

\begin{thebibliography}{10}

\bibitem{dehak_2011_1}
Najim Dehak, Patrick Kenny, R\'{e}da Dehak, Pierre Dumouchel, and Pierre
  Ouellet,
\newblock ``Front-end factor analysis for speaker verification,''
\newblock {\em IEEE Transactions on Audio, Speech and Language Processing},
  vol. 19, no. 4, pp. 788--798, 2011.

\bibitem{snyder_2018_1}
David Snyder, Daniel Garcia-Romero, Gregory Sell, Daniel Povey, and Sanjeev
  Khudanpur,
\newblock ``X-vectors: Robust {DNN} embeddings for speaker recognition,''
\newblock in {\em IEEE ICASSP}, 2018.

\bibitem{heigold_2016_1}
Georg Heigold, Ignacio Moreno, Samy Bengio, and Noam Shazeer,
\newblock ``End-to-end text-dependent speaker verification,''
\newblock in {\em IEEE ICASSP}, 2016.

\bibitem{wan_2018_1}
Li~Wan, Quan Wang, Alan Papir, and Ignacio~Lopez Moreno,
\newblock ``Generalized end-to-end loss for speaker verification,''
\newblock in {\em IEEE ICASSP}, 2018.

\bibitem{desplanques_2020_1}
Brecht Desplanques, Jenthe Thienpondt, and Kris Demuynck,
\newblock ``{ECAPA-TDNN}: {E}mphasized channel attention, propagation and
  aggregation in {TDNN} based speaker verification,''
\newblock in {\em Interspeech}, 2020.

\bibitem{tan2021_1}
Zhenning Tan, Yuguang Yang, Eunjung Han, and Andreas Stolcke,
\newblock ``Improving speaker identification for shared devices by adapting
  embeddings to speaker subsets,''
\newblock in {\em IEEE Automatic Speech Recognition and Understanding
  Workshop}, 2021.

\bibitem{prince_2007_1}
Simon~J.D. Prince and James~H. Elder,
\newblock ``Probabilistic linear discriminant analysis for inferences about
  identity,''
\newblock in {\em IEEE International Conference on Computer Vision}, 2007.

\bibitem{silnova_2020_1}
Anna Silnova, Niko Brümmer, Johan Rohdin, Themos Stafylakis, and Luk\'{a}\v{s}
  Burget,
\newblock ``Probabilistic embeddings for speaker diarization,''
\newblock in {\em Odyssey}, 2020.

\bibitem{pelecanos2021_1}
Jason Pelecanos, Quan Wang, and Ignacio Lopez~Moreno,
\newblock ``Dr-vectors: {D}ecision residual networks and an improved loss for
  speaker recognition,''
\newblock in {\em Interspeech}, 2021, pp. 4603--4607.

\bibitem{vaswani2017_1}
Ashish Vaswani, Noam Shazeer, Niki Parmar, Jakob Uszkoreit, Llion Jones,
  Aidan~N. Gomez, \L{}ukasz Kaiser, and Illia Polosukhin,
\newblock ``Attention is all you need,''
\newblock in {\em Proceedings of the 31st International Conference on Neural
  Information Processing Systems}, 2017, p. 6000–6010.

\bibitem{tong_2020_1}
Ying Tong, Wei Xue, Shanluo Huang, Lu~Fan, Chao Zhang, Guohong Ding, and
  Xiaodong He,
\newblock ``The {JD AI} speaker verification system for the {FFSVC} 2020
  challenge,''
\newblock in {\em Interspeech}, 2020.

\bibitem{safari_2020_1}
Pooyan Safari, Miquel India, and Javier Hernando,
\newblock ``Self-attention encoding and pooling for speaker recognition,''
\newblock in {\em Interspeech}, 2020.

\bibitem{mary_2020_1}
N~J Metilda~Sagaya Mary, S~Umesh, and Sandesh~V Katta,
\newblock ``{S-vectors and TESA}: {S}peaker embeddings and a speaker
  authenticator based on transformer encoder,''
\newblock {\em IEEE Transactions on Audio, Speech and Language Processing},
  2020.

\bibitem{wang_2022_1}
Rui Wang, Junyi Ao, Long Zhou, Shujie Liu, Zhihua Wei, Tom Ko, Qing Li, and
  Yu~Zhang,
\newblock ``Multi-view self-attention based transformer for speaker
  recognition,''
\newblock in {\em IEEE ICASSP}, 2022.

\bibitem{Li2020TextIndependentSV}
Jingyu Li and Tan Lee,
\newblock ``Text-independent speaker verification with dual attention
  network,''
\newblock in {\em Interspeech}, 2020.

\bibitem{Jung2021graphatt}
Jee-weon Jung, Hee-Soo Heo, Ha-Jin Yu, and Joon~Son Chung,
\newblock ``Graph attention networks for speaker verification,''
\newblock in {\em IEEE ICASSP}, 2021, pp. 6149--6153.

\bibitem{wang2020version}
Quan Wang and Ignacio Lopez~Moreno,
\newblock ``Version control of speaker recognition systems,''
\newblock {\em arXiv:2007.12069}, 2020.

\bibitem{shen2019lingvo}
Jonathan Shen, Patrick Nguyen, Yonghui Wu, Zhifeng Chen, Mia~X Chen, Ye~Jia,
  Anjuli Kannan, Tara Sainath, Yuan Cao, Chung-Cheng Chiu, et~al.,
\newblock ``Lingvo: {A} modular and scalable framework for sequence-to-sequence
  modeling,''
\newblock {\em arXiv:1902.08295}, 2019.

\bibitem{ba2016_1}
Jimmy~Lei Ba, Jamie~Ryan Kiros, and Geoffrey~E. Hinton,
\newblock ``Layer normalization,''
\newblock in {\em Advances in Neural Information Processing Systems 2016 Deep
  Learning Symposium}, 2016.

\bibitem{Hinton2015DistillingTK}
Geoffrey Hinton, Oriol Vinyals, and Jeff Dean,
\newblock ``Distilling the knowledge in a neural network,''
\newblock {\em arXiv:1503.02531}, 2015.

\bibitem{gulati2020conformer}
Anmol Gulati, James Qin, Chung-Cheng Chiu, Niki Parmar, Yu~Zhang, Jiahui Yu,
  Wei Han, Shibo Wang, Zhengdong Zhang, Yonghui Wu, and Ruoming Pang,
\newblock ``Conformer: Convolution-augmented transformer for speech
  recognition,''
\newblock in {\em Interspeech}, 2020.

\bibitem{prabhavalkar2015automatic}
Rohit Prabhavalkar, Raziel Alvarez, Carolina Parada, Preetum Nakkiran, and
  Tara~N Sainath,
\newblock ``Automatic gain control and multi-style training for robust
  small-footprint keyword spotting with deep neural networks,''
\newblock in {\em IEEE ICASSP}, 2015, pp. 4704--4708.

\bibitem{wang2022attentive}
Quan Wang, Yang Yu, Jason Pelecanos, Yiling Huang, and Ignacio~Lopez Moreno,
\newblock ``Attentive temporal pooling for conformer-based streaming language
  identification in long-form speech,''
\newblock {\em arXiv preprint arXiv:2202.12163}, 2022.

\bibitem{Chowdhury2018_1}
F~A Rezaur~Rahman Chowdhury, Quan Wang, Ignacio~Lopez Moreno, and Li~Wan,
\newblock ``Attention-based models for text-dependent speaker verification,''
\newblock in {\em IEEE ICASSP}, 2018, pp. 5359--5363.

\bibitem{zhu2018_1}
Yingke Zhu, Tom Ko, David Snyder, Brian Mak, and Daniel Povey,
\newblock ``Self-attentive speaker embeddings for text-independent speaker
  verification,''
\newblock in {\em Interspeech}, 2018, pp. 3573--3577.

\bibitem{nair2010_1}
Vinod Nair and Geoffrey~E. Hinton,
\newblock ``Rectified linear units improve restricted boltzmann machines,''
\newblock in {\em Proceedings of the 27th International Conference on Machine
  Learning}, 2010, pp. 807--814.

\bibitem{Fukushima1980_1}
Kunihiko Fukushima,
\newblock ``Neocognitron: {A} self-organizing neural network model for a
  mechanism of pattern recognition unaffected by shift in position,''
\newblock {\em Biological Cybernetics}, vol. 36, pp. 193--202, 1980.

\bibitem{Kingma2015_1}
Diederik~P. Kingma and Jimmy~Lei Ba,
\newblock ``Adam: A method for stochastic optimization,''
\newblock {\em International Conference on Learning Representations (ICLR)},
  2015.

\bibitem{librivox_2020_1}
LibriVox,
\newblock ``{L}ibri{V}ox: {F}ree public domain audiobooks,''
  https://librivox.org/, 2020.

\bibitem{fan_2020_1}
Y.~Fan, J.W. Kang, L.T. Li, K.C. Li, H.L. Chen, S.T. Cheng, P.Y. Zhang, Z.Y.
  Zhou, Y.Q. Cai, and D.~Wang,
\newblock ``{CN-Celeb: A} challenging {C}hinese speaker recognition dataset,''
\newblock in {\em IEEE ICASSP}, 2020.

\bibitem{Cieri2004_1}
Christopher Cieri, David Graff, Owen Kimball, Dave Miller, and Kevin Walker,
\newblock ``Fisher {E}nglish training speech parts 1 and 2, {LDC2004S13,
  LDC2005S13},''
\newblock {\em Web Download. Philadelphia: Linguistic Data Consortium}, 2004.

\bibitem{Brandschain2020_1}
Linda Brandschain, Kevin Walker, David Graff, Christopher Cieri, Abby Neely,
  Nikki Mirghafori, Barbara Peskin, Jack Godfrey, Stephanie Strassel, Fred
  Goodman, George~R. Doddington, and Mike King,
\newblock ``Mixer 4 and 5 speech {LDC2020S03},''
\newblock {\em Web Download. Philadelphia: Linguistic Data Consortium}, 2020.

\bibitem{garofolo_1993_1}
John~S. Garofolo, Lori~F. Lamel, William~M. Fisher, Jonathan~G. Fiscus,
  David~S. Pallett, Nancy~L. Dahlgren, and Victor Zue,
\newblock ``{TIMIT} acoustic-phonetic continuous speech corpus {LDC93S1},''
\newblock {\em Web Download. Philadelphia: Linguistic Data Consortium}, 1993.

\bibitem{lippmann1987multi}
Richard~P. Lippmann, Edward~A. Martin, and Douglas~B. Paul,
\newblock ``Multi-style training for robust isolated-word speech recognition,''
\newblock in {\em IEEE ICASSP}, 1987, vol.~12, pp. 705--708.

\bibitem{ko2017study}
Tom Ko, Vijayaditya Peddinti, Daniel Povey, Michael~L. Seltzer, and Sanjeev
  Khudanpur,
\newblock ``A study on data augmentation of reverberant speech for robust
  speech recognition,''
\newblock in {\em IEEE ICASSP}, 2017, pp. 5220--5224.

\bibitem{kim2017generation}
Chanwoo Kim, Ananya Misra, Kean Chin, Thad Hughes, Arun Narayanan, Tara
  Sainath, and Michiel Bacchiani,
\newblock ``Generation of large-scale simulated utterances in virtual rooms to
  train deep-neural networks for far-field speech recognition in {Google
  Home},''
\newblock in {\em Interspeech}, 2017.

\bibitem{garcia-romero_2012_1}
Daniel Garcia-Romero, Xinhui Zhou, and Carol~Y. Espy-Wilson,
\newblock ``Multicondition training of {G}aussian {PLDA} models in i-vector
  space for noise and reverberation robust speaker recognition,''
\newblock in {\em IEEE ICASSP}, 2012, pp. 4257--4260.

\bibitem{lei_2012_1}
Yun Lei, Lukas Burget, Luciana Ferrer, Martin Graciarena, and Nicolas Scheffer,
\newblock ``Towards noise-robust speaker recognition using probabilistic linear
  discriminant analysis,''
\newblock in {\em IEEE ICASSP}, 2012, pp. 4253--4256.

\bibitem{avila_2014_1}
Anderson~R. Avila, Milton Sarria-Paja, Francisco~J. Fraga, Douglas
  O'Shaughnessy, and Tiago~H. Falk,
\newblock ``Improving the performance of far-field speaker verification using
  multi-condition training: The case of {GMM-UBM} and i-vector systems,''
\newblock in {\em Interspeech}, 2014, pp. 1096--1100.

\bibitem{huang_2019_1}
Chien-Lin Huang,
\newblock ``Exploring effective data augmentation with {TDNN-LSTM} neural
  network embedding for speaker recognition,''
\newblock in {\em IEEE Automatic Speech Recognition and Understanding Workshop
  (ASRU)}, 2019.

\end{thebibliography}
\end{document}